\mag\magstephalf
\documentclass[12pt]{report}
\usepackage{graphicx}
\usepackage[T2A]{fontenc}
\usepackage[cp1251]{inputenc}
\usepackage[english]{babel}
\usepackage{amsmath,amssymb}

\font\blarge=cmr17 scaled\magstep1

\def\bfrac#1#2#3{\mathchoice%
        {\mathop{\kern0pt\vphantom{\sum}\lower4.1pt\hbox{$\fam0#1\textfont0=\blarge$}}\limits_{#2}^{#3}}%
        {\hbox{$\mathop{\kern0pt\fam0#1}\limits_{#2}^{#3}$}}%
        {\hbox{$\mathop{\kern0pt\fam0#1}\limits_{#2}^{#3}$}}%
        {\hbox{$\mathop{\kern0pt\fam0#1}\limits_{#2}^{#3}$}}}

\begin{document}

\centerline{{\bf NONEQUATORIAL CIRCULAR ORBITS of SPINNING PARTICLES}}
\centerline{{\bf in the SCHWARZSCHILD-de SITTER BACKGROUND}}
\vspace{0.5cm}
\centerline{{\bf Roman Plyatsko, Volodymyr Panat, Mykola Fenyk}}

\vspace{0.4cm}
\centerline{Pidstryhach Institute for Applied Problems in Mechanics and Mathematics,}
\centerline{National Academy of Sciences of Ukraine}
\centerline{3-b Naukova Street, Lviv, 79060, Ukraine}
\centerline{E-mail: plyatskor@gmail.com}
 
\vspace{5mm}
{\bf Abstract} In this paper analytical solutions of the Mathisson-Papapetrou equations that describe nonequatorial circular orbits of a spinning particle in the Schwarz- schild-de Sitter background are studied, and the role of the cosmological constant is emphasized. It is shown that generally speaking a highly relativistic velocity of the particle is a necessary condition of motion along this type of orbits, with an exception of orbits locating close to the position of the static equilibrium, where low velocities are possible as well. Depending on the correlation between the spin orientation of the particle and its orbital velocity some of the possible nonequatorial circular orbits exist due to the repulsive action on the particle caused by  the spin-gravity coupling and the others are caused by the attractive action. 
Here  values of the  energy of the particle on the corresponding orbits are also analyzed.

\vspace{3mm}

\vspace{3mm}

\vspace{5mm}

\centerline{{\bf 1. Introduction}}
\vspace{3mm}
Description of circular orbits is considered very important when it comes to investigating different possible types of test particles motions in gravitational fields. For simple test particles which follow geodesic lines in Schwarzschild's and Kerr's metrics the corresponding analysis is presented, for example, in classical books \cite{Misner,Chandra}. Despite the fact that circular orbits describe only few partial cases of all possible trajectories, they illustrate some important features of influence that a strong gravitational field has on a fast moving particle. In particular, these orbits are interesting in the context of studying possible synchrotron radiation \cite{Bre,Chrz,Pl05}.

Highly relativistic equatorial circular orbits of test particles with inner rotation (spin) in the Schwarzschild, Kerr, and Schwarzschild--de Sitter backgrounds are thoroughly investigated \cite{Pl10,Pl12,Pl13,Pl15,Pl17} using  the Mathisson--Papapetrou (MP) equations \cite{Mathis,Papa} which were derived for classical (nonquantum) spinning particles and proved to follow from the general relativistic Dirac equation in certain classical approximation later on \cite{Dirac}.

As far as types of orbits of the test particle in Schwarzshild's and Kerr's background are concerned, the majority of researchers turn their attention to equatorial orbits only, and only few of them are looking into the possibility of existence of nonequatorial orbits as well. For example, in \cite{Bon} it was suggested that such orbits might be allowed in the Kerr metric according to the geodesic equations.  
However, in \cite{Fel} with more careful analysis being conducted, this suggestion was refuted and it was indubitably proved that the correspondent solutions do not exist. In the case of a spinning test particle its nonequatorial circular orbits in Schwarzschild's and Kerr's backgrounds are considered in \cite{Pl15,Pl82,Pl88}. 

Nowadays, a lot of attention is drawn to unraveling  the role which the cosmological constant $\Lambda$ plays in astrophysics and cosmology, so it would be interesting to study the influence of $\Lambda$ on motion of a spinning particle in the Schwarz\-schild--de Sitter background. In this context we point out the recent papers \cite{Mort,Kunst}. Naturally, it is important to continue further research in line with the investigation presented in \cite{Pl17,Mort,Kunst}.The purpose of this paper is to investigate the impact of $\Lambda$ in the case of possible nonequatorial circular orbits of the spinning particle in the Schwarzschild--de Sitter background. In more general context the data concerning influence of $\Lambda$ on highly relativistic spinning particles are useful for more detailed analysis of formation stages of the Universe.

The paper is organized in the following way. Basic information about the MP equations is presented in Sect. 2. Sect. 3  is devoted to the analysis of the main relations following from the MP equations for possible nonequatorial circular orbits of the spinning particle in the Schwarzschild--de Sitter background. Thoroughgoing analysis of different special cases of these orbits is given in Sect. 4. In Sect. 5 the energy of the spinning particle on the nonequatorial circular orbits is estimated. We conclude in Sect. 6. 

\vspace{3mm}
\centerline{{\bf 2. Initial equations}}
\vspace{3mm}
We use the MP equations in the form that was presented in \cite{Mathis}:
\begin{equation}\label{1}
\frac D {ds} \left(mu^\lambda + u_\mu\frac {DS^{\lambda\mu}}
{ds}\right)= -\frac {1} {2} u^\pi S^{\rho\sigma}
R^{\lambda}_{~\pi\rho\sigma},
\end{equation}
\begin{equation}\label{2}
\frac {DS^{\mu\nu}} {ds} + u^\mu u_\sigma \frac {DS^{\nu\sigma}}
{ds} - u^\nu u_\sigma \frac {DS^{\mu\sigma}} {ds} = 0,
\end{equation}
\begin{equation}\label{3}
S^{\lambda\nu} u_\nu = 0
\end{equation}
where $u^\lambda\equiv dx^\lambda/ds$ is the particle's 4-velocity,
$S^{\mu\nu}$ is the antisymmetric tensor of spin, $m$ and $D/ds$ are the mass and the covariant derivative along $u^\lambda$, respectively. Here, and in the following, greek indices run through 1, 2, 3, 4 and latin indices run through 1, 2, 3; the signature of the metric (--, --, --, +) and the unites $c=G=1$ are chosen.

Later instead of (\ref{3}) other relations for the tensor of spin were introduced and now these relations are known as supplementary conditions for the MP equations. 
Different  conditions are connected with different representative points that describe the motion of the particle \cite{Papa}. As in our previous papers \cite{Pl12,Pl15,Pl17,Pl11,Pl16a,Pl16b}, below we use condition (\ref{3}). Detailed analysis of physical meaning of this condition is
carried out in \cite{Costa} and, what is most important,  the known helical solutions of Eqs. (\ref{1})--(\ref{3}) are explained. Other aspects of using condition (\ref{3}) in practical calculations are discussed in \cite{Costa-2,Kyr}.

Working with Eqs. (\ref{1})--(\ref{3}) it is convenient to use their representation through the spin 3-vector $S_i$ which is defined by
\begin{equation}\label{10}
S_i=\frac{1}{2} \sqrt{-g} \varepsilon_{ikl}  S^{kl},
\end{equation}
where $\varepsilon_{ikl}$ is the spatial Levi-Civita symbol.
For example, with condition (\ref{3}) the three independent equations of set (\ref{2}) can be rewritten as
$$
u_4 \dot S_i - \dot u_4 S_i +  2(\dot u_{[4} u_{i]} -
u^\pi u_\rho \Gamma^\rho_{\pi[4} u_{i]})S_k u^k
$$
\begin{equation}\label{11}
+ 2S_n \Gamma^n _{\pi [4} u_{i]} u^\pi =0,
\end{equation}
where $\Gamma^\rho_{\pi\sigma}$ are the Christoffel symbols, a dot denotes the usual differentiation with respect to the proper time $s$, and square brackets denote antisymmetrization of indices. Equation (\ref{11}) can be obtained directly from the three equations of set (\ref{2}) with the spatial indices by using the expression
\begin{equation}\label{11a}
 S^{i4}=\frac{u_k}{u_4}S^{ki},
\end{equation}
which follows from (\ref{3}), and the relation
\begin{equation}\label{11b}
S^{kl}=\frac{1}{\sqrt{-g}}\varepsilon^{klm}S_m
\end{equation}
that is the inverse relation to (\ref{10}).

\vspace{4mm}
\centerline{{\bf 3. Relations which follow from the MP equations}}
\centerline{{\bf in the Schwarzschild--de Sitter metric }}
\centerline{{\bf for nonequatorial circular orbits}}

\vspace{3mm}
We use the Schwarzschild-de Sitter metric in the standard coordinates
 $x^1=r, \quad x^2=\theta, \quad x^3=\varphi, \quad x^4=t$. 
Then the nonzero components of the metric tensor $g_{\mu\nu}$ are
\[
g_{11}=-\left(1-\frac{2M}{r} - \frac{\Lambda r^2}{3}\right)^{-1} , \quad g_{22}=- r^2,
\]
\begin{equation}\label{12}
g_{33}=-r^2\sin^2\theta, \quad
g_{44}=1-\frac{2M}{r} - \frac{\Lambda r^2}{3},
\end{equation}
where $M$ and $\Lambda>0$ are, respectively, the mass parameter and the cosmological constant. 

Let us check if Eqs. (\ref{1})--(\ref{3}) in metric (\ref{12}) have solutions that describe nonequatorial circular orbits with
\begin{equation}
\label{6.1}
r=const\ne 0, \quad \theta=const\ne 0, \pi/2, \pi,
\end{equation}
\begin{equation}
\label{6.2}
u^3=\frac{d\varphi}{ds}=const\ne 0, \quad u^4=\frac{dt}{ds}=const\ne 0.
\end{equation}
Taking into account Eq. (\ref{2}) in the form (\ref{11}) by (\ref{6.1}) we
obtain
\begin{equation}
\label{6.3}
\dot S_1+S_3u^3u^4u_4(\Gamma^4_{14}-\Gamma^3_{13})=0,
\end{equation}
\begin{equation}
\label{6.4}
\dot S_2-S_3u^3u^4u_4\Gamma^3_{23}=0,
\end{equation}
\begin{equation}
\label{6.5}
\dot S_3+S_1u_3(g^{44}\Gamma^1_{44}-g^{33}\Gamma^1_{33})-
S_2\Gamma^2_{33}u_3g^{33}=0.
\end{equation}
It is easy to see that Eqs. (\ref{6.3})--(\ref{6.5}) have solutions
\begin{equation}
\label{6.6}
S_3\equiv S_{\varphi}=0,
\end{equation}
\begin{equation}
\label{6.7}
S_1=const, \quad S_2=const,
\end{equation}
\begin{equation}
\label{6.8}
S_1(g^{44}\Gamma^1_{44}-g^{33}\Gamma^1_{33})-S_2g^{33}\Gamma^2_{33}=0.
\end{equation}
After calculating the Christoffel symbols $\Gamma^1_{33}$, $\Gamma^1_{44}$, and $\Gamma^2_{33}$ for metric (\ref{12}), relation (\ref{6.8}) becomes
\begin{equation}\label{6.9}
S_1\left(1-\frac{3M}{r}\right)+S_2\frac{\cos\theta}{r\sin\theta}=0.
\end{equation}

Concerning the set of Eqs. (\ref{1}) we note that with relation (\ref{6.1}), (\ref{6.2}), (\ref{6.6}), and  (\ref{6.7}) the two equations with
$\lambda=3$ and  $\lambda=4$ are satisfied automatically. The other two  with $\lambda=1$ and  $\lambda=2$ can be rewritten through the 3-vector components  $S_1$ and $S_2$ as
$$
m(\Gamma^1_{33}u^3u^3+\Gamma^1_{44}u^4u^4)+u^3(\Gamma^1_{33}-
g^{44}g_{33}\Gamma^1_{44})
$$
\begin{equation}
\label{6.10}
\times\frac{1}{\sqrt{-g}}(g_{44}\Gamma^4_{14}u^4u^4S_2
+g_{33}\Gamma^3_{13}u^3u^3S_2-g_{33}\Gamma^3_{23}u^3u^3S_1)
=-\frac{3M}{r^3}u^3S_2\sin\theta,
\end{equation}
$$
m\Gamma^2_{33}u^3u^3+
u^3\Gamma^2_{33}\frac{1}{\sqrt{-g}}(g_{44}\Gamma^4_{14}u^4u^4S_2
+g_{33}\Gamma^3_{13}u^3u^3S_2
$$
\begin{equation}
\label{6.11}
-g_{33}\Gamma^3_{23}u^3u^3S_1)=-\frac{3M}{r^3}u^3S_1\sin\theta.
\end{equation}
In addition to (\ref{6.10}) and (\ref{6.11}), the components $u^3$ and $u^4$ of the 4-velocity have to satisfy the relation
\begin{equation}
\label{6.20}
g_{33}(u^3)^2+g_{44}(u^4)^2=1
\end{equation}
as a partial case of more general expression $u_\mu u^\mu =1$.
Taking into account explicit expressions for the Christoffel symbols and relation (\ref{6.9}) we obtain from (\ref{6.10}) and (\ref{6.11})
\begin{equation}
\label{6.23}
u^3=-\frac{mr}{6S_2\sin\theta}\left(1-\frac{\Lambda r^3}{3M}\right) \quad (S_2\ne 0)
\end{equation}
(at $S_2=0$ Eqs. (\ref{6.10}) and (\ref{6.11}) do not have solutions). 
By (\ref{6.20}) and (\ref{6.23}) we write
\begin{equation}
\label{6.24}
u^4=\frac{mr^2}{6|S_2|}\left(1-\frac{2M}{r}-\frac{\Lambda r^2}{3}\right)^{-1/2} 
 \left[\left(1- \frac{\Lambda r^3}{3M}\right)^2 +
\frac{36S_2^2}{m^2r^4}\right]^{1/2}
\end{equation}
(we consider direct motions in time when $u^4>0$ and the condition
$g_{44}>0$ is used).

Using expressions  (\ref{6.23}) and  (\ref{6.24}) we
find the necessary and sufficient condition for compatibility of  Eqs.
 (\ref{6.10})--(\ref{6.20}) in the form
$$
\sin^2\theta=\left(1-\frac{2M}{r}-\frac{\Lambda r^2}{3}\right)\left(1-\frac{\Lambda r^3}{3M}\right) 
\left[\frac Mr \left(4-\frac{9M}{r}-\frac{\Lambda r^3}{3M}\right)\right.
$$
\begin{equation}
\label{6.26}
\left.\times\left(1-\frac{\Lambda r^3}{3M}+\frac{36S^2_2}{m^2r^4}
\frac{1}{1-\frac{\Lambda r^3}{3M}}\right)
-6\left(1-\frac{2M}{r}-\frac{\Lambda r^2}{3}\right)
\left(1-\frac{3M}{r}\right)\right]^{-1}.
\end{equation}

In the written below  it is necessary to take into account the physical condition for the spinning test particle \cite{Wald}
\begin{equation}\label{6.27}
 \varepsilon \equiv \frac{|S_0|}{mr}\ll 1,
\end{equation}
where $|S_0|$ is the absolute value of its spin. By contracting Eq. (\ref{3}) with $S_{\mu\nu}$ it is shown that $S_0$ is a constant of motion of the MP Eqs. (\ref{1})--(\ref{3}) and by definition
\begin{equation}\label{6.27a}
 S_0^2=\frac12 S_{\mu\nu}S^{\mu\nu}.
\end{equation}
It is not difficult to calculate that for the  solutions of the MP equations for the nonequatorial circular orbits in the Schwarzschild--de Sitter background the relation 
\begin{equation}\label{6.27b}
S_0^2=\frac{S_2^2}{r^2(u_4)^2}\left[1+\left(1-\frac{2M}{r}-\frac{\Lambda r^2}{3}\right)\cos^2\theta
 \left(1-\frac{3M}{r}\right)^{-2}\sin^{-2}\theta\right]
\end{equation}
takes place.
Using Eq. (\ref{6.24}) from (\ref{6.27b}) we get
\begin{equation}\label{6.28}
\frac{|S_0|}{mr}=\frac{6S^2_2}{m^2r^4}\frac{A}{B},
\end{equation}
where
$$
A=\left[\left|-6\left(1-\frac{2M}{r}-\frac{\Lambda r^2}{3}\right)
\left(1-\frac{3M}{r}\right)\right.\right.
$$
$$
\times\left.\left.\left(1-\frac{\Lambda r^3}{3M}\right)
+36\frac{S^2_2}{m^2r^4}\left(\frac{4M}{r}-\frac{9M^2}{r^2}-\frac{\Lambda r^2}{3}\right)
\right| \right]^{1/2},
$$
$$
B=\left|\left(1-\frac{3M}{r}\right)\left(1-\frac{\Lambda r^3}{3M}\right)\right|\left(1-\frac{2M}{r}-\frac{\Lambda r^2}{3}\right)^{1/2}
$$
\begin{equation}
\label{6.28a}
\times\left[\left(1-\frac{\Lambda r^3}{3M}\right)^2 +36\frac{S^2_2}{m^2r^4}\right]^{1/2}.
\end{equation}
It follows from (\ref{6.28}) that condition (\ref{6.27}) is satisfied when 
\begin{equation}
\label{6.29}
\frac{6S^2_2}{m^2r^4}\ll 1.
\end{equation}
Then by (\ref{6.27}) and (\ref{6.28})--(\ref{6.29}) the following relation between the left-hand side of (\ref{6.29}) and $\varepsilon$ takes place:
\begin{equation}
\label{6.29a}
\frac{6S^2_2}{m^2r^4} = \frac{\varepsilon}{\sqrt{6}}\left|1-\frac{3M}{r}\right|^{1/2}\left|1-\frac{\Lambda r^3}{3M}\right|^{3/2}(1+O(\varepsilon)).
\end{equation}

Note that expressions (\ref{6.23}), (\ref{6.24}), (\ref{6.26}), (\ref{6.27b})
and (\ref{6.28a}) are exact, in contrast to (\ref{6.29a}) and other relations
below in the next section where we present results of the analytical calculations at condition (\ref{6.29}). This approach is possible due to the existence the natural physical small value $\varepsilon$ which is determined in (\ref{6.27}).

\vspace{4mm}
\centerline{{\bf 4. Conditions of existence of nonequatorial circular orbits}}
\centerline{{\bf of a spinning particle  in the Schwarzschild--de Sitter background}}

\vspace{3mm}
 Below in this section we consider the possibility of the dynamic equilibrium of the spinning particle in the Schwarzschild--de Sitter background in the form of the nonequatorial circular orbits.
From Eqs. (\ref{6.23}), (\ref{6.26}), and (\ref{6.29a}) we obtain the explicit conditions  on the motion of the particle along possible nonequatorial circular orbits in terms of its coordinates and velocity. 

\vspace{3mm}
\centerline{{\bf 4.1} The case $\frac{\Lambda r^3}{3M}\ll 1$ }

\vspace{3mm}
We begin with the partial case
\begin{equation}
\label{6.30-1}
\frac{\Lambda r^3}{3M}\ll 1,
\end{equation}
when the basic expressions for the Schwarzschild--de Sitter background is close to the corresponding expressions for the Schwarzschild background, which is interesting from the physical point of view. Indeed, under conditions $g_{44}>0$ and $\Lambda>0$ we have $M<\frac{r}{2}$
and then it follows from (\ref{6.30-1}) that
\begin{equation}
\label{6.30}
\frac{\Lambda r^2}{3}\ll 1. 
\end{equation}
Note that according to Eq.  (\ref{12}) the only difference between   the Schwarzschild--de Sitter and the  Schwarzschild metrics is the term $\frac{\Lambda r^2}{3}$. So,  (\ref{6.30-1}) let us estimate the small corrections caused by $\Lambda \ne 0$ and check the boundary transition.

Condition (\ref{6.26}) can be satisfied only if the right-hand side of (\ref{6.26}) is within in the interval (0, 1). We analyze Eq. (\ref{6.26}) in the linear approximation in $\Lambda$. 
Then according to (\ref{6.26}) we write
$$
\sin^2\theta=\left[1-\frac{2M}{r}+\frac{\Lambda r^2}{3M}(M-r)\right]
$$
\begin{equation}
\label{6.31}
\times\left[(1+\alpha)\frac Mr \left(4-\frac{9M}{r}\right)+\frac{\Lambda r^2}{3}
\left(1-\frac{9M}{r}\right)-6\left(1-\frac{2M}{r}\right)
\left(1-\frac{3M}{r}\right)\right]^{-1},
\end{equation}
where by (\ref{6.29}) and (\ref{6.29a})
\begin{equation}
\label{6.32}
\alpha\equiv \frac{36S_2^2}{m^2r^4}\ll 1,
\end{equation}
i.e. we put that not only the value in the left-hand side of (\ref{6.29}) but the six times bigger value $\alpha$ in (\ref{6.32}) is also much smaller then 1. Let us estimate when the right-hand side of (\ref{6.31}) has values in the 
interval (0, 1). First of all, we note that for $\Lambda=0$ the right-hand side of (\ref{6.31}) is equal to 1 when the second order algebraic equation for $M/r$ is satisfied:
\begin{equation}
\label{6.32-1}
\left(1-\frac{2M}{r}\right)=(1+\alpha)\frac{M}{r}\left(4-\frac{9M}{r}\right) - 6\left(1-\frac{2M}{r}\right)\left(1-\frac{3M}{r}\right).
\end{equation}
This equation has  real roots which in the linear for  approximation for $\alpha$ are
\begin{equation}
\label{6.32-2}
\left(\frac{M}{r}\right)_1=\frac{7}{15}\left(1-\frac{\alpha}{5}\right), \quad
\left(\frac{M}{r}\right)_2=\frac{1}{3}\left(1-\alpha\right).
\end{equation}
Then according to (\ref{6.31}) the condition $\sin^2\theta<1$ is satisfied if and only if 
\begin{equation}
\label{6.33}
\frac{15}{7} M\left(1+\frac{\alpha}{5}\right)<r<
3M(1+\alpha),
\end{equation}
or approximately, when the small value $\alpha$ is neglected in comparison with 1,  
\begin{equation}
\label{6.34}
\frac{15}{7} M<r<
3M.
\end{equation}
In the corresponding approximation it follows from  the right-hand side of (\ref{6.31}) that the minimal value of $\sin^2\theta$ is equal to 0.465 and 
is achieved at 
$$
r=\frac{5\sqrt 5}{3\sqrt 5 -1}= 2.35M.
$$
At $r=2.25M$ and $r=2.5M$ from (\ref{6.31}) we have $\sin^2\theta=0.5$.

When $\Lambda\ne 0$ it is easy to obtain from (\ref{6.31}) the corresponding corrections to expressions (\ref{6.32-2}) in the linear approximation in $\Lambda$, and a generalization of condition (\ref{6.33}) for a nonzero $\Lambda$ is
\begin{equation}
\label{6.34a}
\frac{15}{7} M\left(1+\frac{\alpha}{5} +\frac{15^2}{7^4}12\Lambda M^2\right)<r<
3M(1+\alpha),
\end{equation}
here the relation $\Lambda M^2\ll 1$ takes place as a simple consequence of (\ref{6.30}) for $\Lambda>0$ and $g_{44}>0$. 

So, relations (\ref{6.31}) and (\ref{6.33}) determine the space region of existence of the nonequatorial circular orbits of the spinning particle in the Schwarzschild--de Sitter background with condition (\ref{6.32}).

For motions on the orbits with an arbirtuary value of $r$ from the interval which is determined by (\ref{6.34a}) and with the corresponding value of $\sin^2\theta$ which follows from (\ref{6.31}) the spinning particle must possess the orbital velocity that is determined by (\ref{6.23}). In the certain sense this situation is similar to the case of the equatorial circular orbits of the spinning particle in the Schwarzschild--de Sitter background  which were considered in \cite{Pl17}. For example, there is correspondence between expression (\ref{6.23}) and the expression for $u^3$  in the case of the equatorial circular orbits  obtained in
\cite{Pl17} for $r<3M$:
\begin{equation}
\label{6.34-3}
u^3 = - \frac{1}{r} \left(\frac{3M}{r} - 1\right)^{-1/2}\left(1-\frac{2M}{r} - \frac{\Lambda r^2}{3}\right)^{1/4}\frac{1}{\sqrt{\varepsilon}}(1+O(\varepsilon)).
\end{equation}
Indeed, according to (\ref{6.9}) for $\sin\theta=1$ and $r\ne 3M$ we have 
$S_1=0$, i.e. the spin is orthogonal to the equatorial plane and by  (\ref{6.34a}) the condition $\sin\theta=1$ is satisfied for
\begin{equation}
\label{6.34-2}
r=\frac{15}{7} M\left(1+\frac{{15}^2}{7^4}12\Lambda M^2\right)
\end{equation}
(this expression is written in the approximation when $\alpha$ is neglected and the same approximation is used in (\ref{6.34-3})).
It is not difficult to check that for $\sin\theta=1$ and $r$ from (\ref{6.34-2})  in the corresponding approximation both expressions  (\ref{6.23}) and (\ref{6.34-3}) have the same value
\begin{equation}
\label{6.34-4}
u^3=-2^{-1/2} 3^{-1/4} 5^{1/4}\left(1-\frac{41}{4}\frac{15^2}{7^4}\Lambda M^2\right)\frac{1}{M\sqrt{\varepsilon}}.
\end{equation}
So, when $\sin\theta \to 1$
and $r$ tends to the value that is determined by the right-hand side of (\ref{6.34-2}), the nonequatorial circular orbits tend to the corresponding equatorial circular orbit. It is a feature which shows the common physical nature of the nonequatorial and equatorial circular orbits. (In this context we note that according to the analysis from Sect. IV.D of paper \cite{Pl17},
Eqs. (70)--(76), the corresponding orbits cannot be interpreted as some partial case of the known helical solutions of Eqs. (\ref{1})--(\ref{3}); this issue will be discussed in more detail in Sect. IV after Eq. (\ref{last2}).)

Let us estimate the orbital velocity of the spinning particle that is necessary for its motion on nonequatorial circular orbits from the point of view of an observer which is at rest relative to the source of the Schwarzschild--de Sitter background. For this observer the particle moving with any 4-velocity $u^\mu$ possess the 
Lorentz factor $\gamma$ which is determined by the expression
\begin{equation}\label{6.34p}
\gamma = \sqrt{u_4 u^4}.
\end{equation}
Relationship  (\ref{6.34a}) follows directly from the definition of the $\gamma$ factor
\begin{equation}\label{6.34b}
\gamma=\frac{1}{\sqrt{1-v^2}},
\end{equation}
where $v^2$ is the second power of the particle's 3-velocity relative to the observer. In the case of diagonal metric (\ref{12}), according to the general expression for the 3-velocity components
$v^i$ we write \cite{Landau}
\begin{equation}\label{6.34c}
v^i=\frac{dx^i}{\sqrt{g_{44}}dt}.
\end{equation}
Then for $v^2$ we have
\begin{equation}\label{6.34d}
v^2=v_i v^i = \gamma_{ik}v^i v^k,
\end{equation}
where $\gamma_{ik}$ is the $3$-space metric tensor, with the following relationship between $\gamma_{ik}$ and $g_{\mu\nu}$ for the diagonal metric: $\gamma_{ik}$=-$g_{ik}$. Equation (\ref{6.34p}) follows from (\ref{6.34b}) with $u_\mu u^\mu=1$. After (\ref{6.24}), (\ref{6.29a})
and (\ref{6.34p}), we have
\begin{equation}
\label{6.34e}
\gamma^2 = 1+ \frac{1}{\varepsilon\sqrt{6}}\left|1-\frac{3M}{r}\right|^{-1/2}\left|1-\frac{\Lambda r^3}{3M}\right|^{1/2}.
\end{equation}
It follows from  (\ref{6.34e}) that when $\frac{\Lambda r^3}{3M}$
is not very close to 1 (in particular, in Schwarzschild's background) the relationship
\begin{equation}
\label{6.34f}
\gamma^2 \gg 1
\end{equation}
takes place, i.e. the corresponding value of the velocity of the particle is highly relativistic. We will use Eq. (\ref{6.34e}) below for other cases of nonequatorial circular orbits.

\vspace{3mm}
\centerline{{\bf 4.2}  Other conditions for $\frac{\Lambda r^3}{3M}$}

\vspace{3mm}

Let us consider relation (\ref{6.26}) at the condition
\begin{equation}
\label{k}
\alpha\left| 1-\frac{\Lambda r^3}{3M} \right|^{-1}\ll \left| 1-\frac{\Lambda r^3}{3M} \right|.
\end{equation}
Because $\alpha\ll 1$ this condition is satisfied both for $\Lambda=0$ and for $\Lambda\ne 0$ if $\frac{\Lambda r^3}{3M}$ is not very close to 1. According to  (\ref{6.23}) and  (\ref{6.32}) it follows from  (\ref{k}) that
$(ru^3)^2\gg 1$ and it means that the particle velocity is highly relativistic.
Then Eq.  (\ref{6.26}) at  (\ref{k}) takes the form
$$
y^2 x^5 (1-\sin^2\theta) + y[x^2 - x^3 - (x^2 -9x)\sin^2\theta]
$$
\begin{equation}
\label{6.36}
+ 1- \frac{2}{x} - \left(\frac{34}{x} -\frac{45}{x^2} -6\right)\sin^2\theta = 0,
\end{equation}
where for the sake of brevity we use the notation
\begin{equation}
\label{6.35}
x=\frac{r}{M}, \quad y=\frac{\Lambda M^2}{3}.
\end{equation}
By (\ref{6.35}) Eq. (\ref{6.36}) is the second order algebraic equation for $\Lambda$, where the corresponding coefficients depend on $M$, $r$, and $\sin^2\theta$.

\vspace{3mm}
\centerline{{\bf 4.2.1} The case $\sin^2\theta=\delta$ }

\vspace{3mm}

First of all, we want to point out that according to Eq. (\ref{6.36}) the nonequatorial circular orbits are allowed for any $x>3$  with a specific value of $\Lambda$ even when the angle $\theta$ is very close to 0 (in contrast to the case considered above with $\Lambda r^2 \ll 1$). Indeed, if
\begin{equation}
\label{6.37}
\sin^2\theta\equiv \delta, \quad  0<\delta\ll 1,
\end{equation}
Eq. (\ref{6.36}) along with the condition  $g_{44}>0$ has a single positive root
\begin{equation}
\label{6.38}
y=\frac{1}{x^3} + \frac{6\delta}{x^4} (x-3).
\end{equation}
According to Eq. (\ref{6.23}), the necessary value of the velocity of the particle $u^3$ on such orbits is
\begin{equation}
\label{6.39}
u^3=-(1-yx^3)\frac{mMx}{6S_2\sqrt{\delta}}.
\end{equation}
Because from (\ref{6.38})
\begin{equation}
\label{6.40}
1-yx^3 = -6\delta\left(1-\frac{3}{x}\right),
\end{equation}
it follows from (\ref{6.39})
\begin{equation}
\label{6.41}
u^3=\frac{mMx}{S_2}\left(1-\frac{3}{x}\right)\sqrt{\delta}.
\end{equation}

In notation (\ref{6.35}) with the expression for $\alpha$ which follows from (\ref{6.29a}) condition (\ref{k}) means
\begin{equation}
\label{k+1}
\varepsilon\sqrt{6}\left|1-\frac{3}{x}\right|^{1/2}\ll |1-yx^3|^{1/2}.
\end{equation}
Then it is easy to check that the value $y$ in (\ref{6.38}) satisfy   (\ref{k+1}) if and only if 
\begin{equation}
\label{k+2}
\varepsilon\ll \sqrt{\delta}.
\end{equation}

Note that for $S_2>0$ and any fixed $x>3$ in (\ref{6.41}),  $u^3>0$ and for $\delta \to 0$ we have $u^3 \to 0$. According to Eqs. (\ref{6.35}) and (\ref{6.38}), the necessary value of $\Lambda$ is determined by 
\begin{equation}
\label{6.42}
\Lambda=\frac{3}{M^2x^3}+ \frac{18\delta}{M^2x^4}(x-3).
\end{equation}
It follows from (\ref{6.42}) that for any $x>3$ the value of $\Lambda$ tends
to 
\begin{equation}
\label{6.43}
\Lambda_0=\frac{3}{M^2x^3} = \frac{3M}{r^3}
\end{equation}
if $\delta \to 0$, where $\Lambda_0$ is the known value of the cosmological constant that provides the static equilibrium position of the spinning particle in 
the Schwarz\-schild--de Sitter background \cite{Stu,Stu-2}. Equations (\ref{6.37})--(\ref{6.41}) describe a partial case of the nonequatorial circular orbits of the spinning particle which prove the possibility of the dynamical equilibrium position (in other words, the effect of hovering). Note that in contrast to the effect of the static equilibrium, which is independent of the value and orientation of the spin of the particle, the dynamical equilibrium significantly depends on its spin. Indeed, according to (\ref{6.9}) and (\ref{6.37}) for sufficiently small $\delta$ the relation 
$$
|S_1| \gg \frac{|S_2|}{r}
$$
takes place, which means that the spin is oriented along the radial direction with the corresponding accuracy.  According to (\ref{6.34p}) and (\ref{6.41}), the velocity of the particle that is necessary for the realization of nonequatorial circular orbits with (\ref{6.37}) and (\ref{6.41})   corresponds to the Lorentz factor
\begin{equation}
\label{6.44}
\gamma^2=1+\delta^2\left(1-\frac{3}{x}\right)^2\frac{m^2 r^4}{S_2^2}.
\end{equation}
Using (\ref{6.29a}) in (\ref{6.44}), we obtain
\begin{equation}
\label{6.45}
\gamma^2=1+\frac{\sqrt{\delta}}{\varepsilon}.
\end{equation}
It follows from (\ref{6.45}) and (\ref{k+2}) that $\gamma^2$ is highly relativistic.

A special situation arises when
\begin{equation}
\label{k+3}
\sqrt{\delta}\ll \varepsilon.
\end{equation}
Indeed, let us consider Eq.  (\ref{6.26}) at the condition
\begin{equation}
\label{k+3a}
|1-yx^3|\ll \frac{\alpha}{|1-yx^3|}.
\end{equation}
Then Eq. (\ref{6.26}) in notation (\ref{6.35}) becomes
$$
\delta\left[\frac{1}{x}\left(4-\frac{9}{x}-yx^3\right)\frac{\alpha}{1-yx^3}- 6\left(1-\frac{2}{x}-yx^2\right)\left(1-\frac{3}{x}\right)\right]
$$
\begin{equation}
\label{k+3b}
=\left(1-\frac{2}{x}-yx^2\right)(1-yx^3),
\end{equation}
where according to (\ref{6.29a}) and (\ref{6.32})
\begin{equation}
\label{k+3c}
\alpha=\varepsilon\sqrt{6}\left|1-\frac{3}{x}\right|^{1/2}|1-yx^3|^{3/2}.
\end{equation}
It easy to check that Eq. (\ref{k+3b}) at condition (\ref{k+3}) is satisfied 
if and only if
\begin{equation}
\label{k+4}
y=\frac{1}{x^3} + \frac{6\delta}{x^3}\left(1-\frac{3}{x}\right),
\end{equation}
in the linear approximation in $\delta$. Note that condition (\ref{k+3a}) is satisfied at (\ref{k+3}), (\ref{k+3c}) and (\ref{k+4}).
After (\ref{k+3}), (\ref{k+4}) and (\ref{6.34e}) 
the value $\gamma^2$ is close to 1 that 
corresponds to a low velocity of the particle, and $\gamma^2\to 1$ with $\sqrt{\delta}\to 0$ for any fixed $\varepsilon$.

We want to emphasize that Eqs. (\ref{6.37})--(\ref{6.42}) are valid for $x>3$, i.e.
$r>3M$. It is easy to check that for $x<3$ Eq.  (\ref{6.36})   with the condition $g_{44}>0$ does not have any positive roots. It means that under condition
(\ref{6.37}) any nonequatorial circular orbits in the region $r<3M$ are impossible. Further analysis shows that if $\sin^2\theta$ is not small in the sense of Eq. (\ref{6.37}), it follows from (\ref{6.36}) that the nonequatorial circular orbits are possible both at $x>3$ and $x<3$. As an example, we consider  the special case when $\sin^2\theta=0.5$ down below.

\vspace{3mm}
\centerline{{\bf 4.2.2} The case $\sin^2\theta=0.5$ }

\vspace{3mm}

First, we note that below we take into account Eq. (\ref{6.29a}) in notation
(\ref{6.35}).
In the case $\sin^2\theta=0.5$ for $x>3$ and $g_{44}>0$ Eq. (\ref{6.36}) has a single positive root
\begin{equation}
\label{6.46}
y=\frac{4x-9}{x^4}.
\end{equation}
According to (\ref{6.23}) the necessary value of $u^3$ is
\begin{equation} 
\label{6.47}
u^3=\frac{mM}{S_2\sqrt{2}}(x-3).
\end{equation}
By (\ref{6.34p}) and (\ref{6.47}) this value corresponds to the Lorentz factor
\begin{equation}
\label{6.48}
\gamma^2= 1+\frac{1}{\varepsilon\sqrt{2}}.
\end{equation}
After (\ref{6.27}) relation (\ref{6.48}) shows that $\gamma^2\gg 1$.

In the region $x<3$  with the condition $g_{44}>0$ Eq. (\ref{6.36}) has two positive roots:
\begin{equation}
\label{6.49}
y_1=\frac{4x-9}{x^4}
\end{equation}
for $2.25<x<3$
and
\begin{equation}
\label{6.50}
y_2=\frac{2x-5}{x^3}
\end{equation}
for $2.5<x<3$. In the case (\ref{6.49}) by (\ref{6.23}) the corresponding value
of $u^3$ is 
\begin{equation}
\label{6.51}
u^3=-\frac{mM}{S_2\sqrt{2}}(3-x),
\end{equation}
whereas for (\ref{6.50}) we have
\begin{equation}
\label{6.52}
u^3=-\frac{mMx\sqrt{2}}{3S_2}(3-x).
\end{equation}
Values (\ref{6.51}) and (\ref{6.52}) correspond to the Lorentz factor
\begin{equation}
\label{6.53}
\gamma^2=1+\frac{1}{\varepsilon\sqrt{2}}
\end{equation}
and 
\begin{equation}
\label{6.54}
\gamma^2=1+\frac{\sqrt{x}}{\varepsilon\sqrt{3}},
\end{equation}
respectively. It is easy to check that in cases (\ref{6.46}), (\ref{6.49}) and
(\ref{6.50}) condition (\ref{k}) is satisfied.

Let us summarize the results obtained in Sects. 3  and 4.
First of all, it is worth noticing  a common feature of all possible nonequatorial circular orbits of the spinning particle in the Schwarzschild--de Sitter background that are described with 
(\ref{6.34f}), (\ref{6.48}), (\ref{6.53}), and (\ref{6.54}). This is that  the  particle must posses highly relativistic velocity with $\gamma^2$ proportional to $1/\varepsilon$ in order for these orbits to be possible, with the only exception for orbits with a very small value of $\sin^2\theta$ when $\gamma^2$ is close to 1. Secondly, there is also essential distinction between different types of these orbits: for the fixed sign of $S_2$ the direction of the orbital rotation which is determined by the sign of $u^3$  in the region $x>3$ (Eqs. (\ref{6.23}),  (\ref{6.51}), and (\ref{6.52})) is opposite to that in the region $x<3$ (Eqs. (\ref{6.41}) and (\ref{6.47}). This situation is similar to the cases of the equatorial circular orbits of the spinning particle in the Schwarzschild--de Sitter background that were studied in \cite{Pl17} where the effects of the strong spin-gravity-$\Lambda$ repulsion for $r<3M$ and the attraction for $r>3M$ were discussed.

It is appropriate to compare the circular orbits of the spinning particle in the Schwarzschild--de Sitter background that were considered in both \cite{Pl17} and the present paper with the circular orbits in the Minkowski spacetime which are a partial case of the known helical orbits. In the coordinates  $x^1=r, \quad x^2=\theta, \quad x^3=\varphi $ for the circular orbits in the plane $\theta=\pi/2$ in the Minkowski spacetime the relation between the angular velocity $u^3=d\varphi/ds$ and the single nonzero component of the particle spin vector $S_2\equiv S_\theta$ is
\begin{equation}
\label{last1}
u^3=\frac{mr}{S_2}.
\end{equation}
This relation is valid for any orbital velocity including its very high values
with $r|u^3|\gg 1$. At the same time for the highly relativistic nonequatorial circular orbits in Schwarzschild's background we have
\begin{equation}
\label{last2}
u^3=-\frac{mr}{6S_2\sin\theta}
\end{equation}
(this relation follows from (\ref{6.23}) for $\Lambda=0$ at $M\ne 0$). Note that for $\sin\theta \to 1$, i.e. when the circular orbits go close to the equatorial plane, relation (\ref{last2}) does not transform into relation (\ref{last1}). This fact is connected with different nature of the two types of the circular orbits for which Eqs. (\ref{last1}) and (\ref{last2}) take place.
In this context we draw the reader's attention to the analysis of the possibility of interpreting the highly relativistic circular orbits as corresponding to the partial case of helical orbits from Sect. IV in \cite{Pl17}. 

\vspace{4mm}
\centerline{{\bf 5. Energy on the nonequatorial circular orbits }}

\vspace{3mm}

Along with its velocity, another  important characteristic of motion of a particle along its orbit is the value of its energy. Because of the symmetry of the Schwarzschild--de Sitter metric, the MP equations demand that the energy of the spinning particle must be their constant of motion.  It is known that equations (\ref{1}) and (\ref{2}) have the Killing vector $\xi^t = \partial /\partial t$ that corresponds to the constant of motion $E$ as the particle's energy \cite{Micol,Tod},
\begin{equation}
\label{6.54a}
E=P_t - \frac{1}{2}g_{t\mu,\nu}S^{\mu\nu},
\end{equation}
where $P_t\equiv P_4$ is the time component of the particle 4-momentum.
There is relation between the momentum $P^\nu$ and the particle 4-velocity $u^\nu$
\begin{equation}
\label{6.54ab}
P^\nu = mu^\nu + u_\lambda\frac{DS^{\nu\lambda}}{ds}.
\end{equation}
It follows from (\ref{6.54a}) and (\ref{6.54ab})  at condition (\ref{3}) 
 in the coordinates chosen above  
$x^1=r, \quad x^2=\theta, \quad x^3=\varphi, \quad x^4=t$ that
\begin{equation}
\label{6.55}
E=mu_4 +g_{44}u_\mu \frac {DS^{4\mu}} {ds} +\frac{1}{2} S^{14}g_{44,1}.
\end{equation}

First of all, with better clarity and comparability of our results in mind let us write the expression of the energy of the spinless particle in the position of the static radius 
\begin{equation}
\label{6.55-1}
r=\left(\frac{3M}{\Lambda}\right)^{1/3}.
\end{equation}
According to (\ref{6.55}) for the spinless particle we have $E=mu_4$. At the condition of the static position $u^1=0$, $u^2=0$, $u^3=0$ taking into account the relation $u_\mu u^\mu=1$ for metric (\ref{12}) we write
\begin{equation}
\label{6.55-2}
u_4=\left(g^{44}\right)^{-1/2}.
\end{equation}
Inserting the value $r$ from (\ref{6.55-1}) into the expression for $g^{44}$ which follows from (\ref{12}) we obtain
\begin{equation}
\label{6.55-3}
g^{44}=\left(1-\frac{3M}{r}\right)^{-1}
\end{equation}
and then
\begin{equation}
\label{d1}
E=m\left(1-\frac{3M}{r}\right)^{1/2}.
\end{equation}

Taking into account (\ref{11}), (\ref{11a}), (\ref{6.9}), and (\ref{6.23}), after direct calculations, from (\ref{6.55}) we obtain the expression of the energy of the spinning particle on the nonequatorial circular orbits that were considered in Sect. 4.
\begin{equation}
\label{6.55a}
E=m\left(1-\frac{2M}{r} - \frac{\Lambda r^2}{3}\right)^{1/2}\left[1+\alpha^{-1}\left(1-\frac{\Lambda r^3}{3M}\right)^{2}\right]^{1/2} +K,
\end{equation}
where
$$
K=\frac{m}{6}\sqrt{\alpha}\left(1-\frac{2M}{r} - \frac{\Lambda r^2}{3}\right)^{1/2}\left(1-\frac{\Lambda r^3}{3M}\right) \left[\alpha+\left(1-\frac{\Lambda r^3}{3M}\right)^2\right]^{-1/2}
$$
$$
\times\left[\frac{M}{r}\left(4-\frac{9M}{r} - \frac{\Lambda r^3}{3M}\right) 
\left(1-\frac{3M}{r}\right)^{-1} \left(1-\frac{2M}{r} - \frac{\Lambda r^2}{3}\right)^{-1}  \right]^{-1},
$$
and $\alpha$ is determined in (\ref{6.32}).

Let us estimate the energy for the case of the orbits with $\sin^2\theta=\delta$. It follows from (\ref{6.55a}) that
\begin{equation}
\label{d1a}
E=m\left(1-\frac{3M}{r}\right)^{1/2} \left(1-\frac{3M\delta}{r}\right)^{2}\left(1+\frac{\sqrt{\delta}}{\varepsilon}\right)^{-1/2}.
\end{equation}
As it can easy be seen,  the right-hand side of (\ref{d1a})  is always        less than that of  (\ref{d1}) 
and tends to it  when $\delta\to 0$ at fixed $\varepsilon$, whereas when $\sqrt{\delta}\gg {\varepsilon}$ it follows from (\ref{d1a}) that
\begin{equation}
\label{d1b}
E=m\left(1-\frac{3M}{r}\right)^{1/2} \left(1-\frac{3M\delta}{r}\right)^{2}\varepsilon^{1/2}\delta^{-1/4}.
\end{equation}
Note that according to (\ref{6.45}) the case of $\sqrt{\delta}\gg {\varepsilon}$ (with $\delta\ll 1$) corresponds to the highly relativistic circular orbits with $\gamma^2\gg 1$ and  energy (\ref{d1b})  proportional to $\gamma^{-1}$.

In the case of $\sin^2\theta=0.5$ with $r>3M$ it follows from (\ref{6.55a}) that the value of the energy which corresponds to (\ref{6.46}) is
\begin{equation}
\label{6.57}
E=-2^{1/4}\left(1-\frac{3}{x}\right)m\sqrt{\varepsilon}.
\end{equation}
Whereas for $\sin^2\theta=0.5$ with $r<3M$ from (\ref{6.55a}) we find  the values of the energy that correspond to (\ref{6.49}) and (\ref{6.50}) in the form
\begin{equation}
\label{6.58}
E=2^{1/4}\left(\frac{3}{x}-1\right)m\sqrt{\varepsilon}
\end{equation}
and
\begin{equation}
\label{6.59}
E=2\left(\frac{x}{3}\right)^{3/4}\left(\frac{3}{x}-1\right)^{3/2}m\sqrt{\varepsilon},
\end{equation}
respectively. According to (\ref{6.48}), (\ref{6.53}) and (\ref{6.54}) all above-mentioned  values of the energy are proportional to $\gamma^{-1}$.

\vspace{4mm}
\centerline{{\bf 6. Conclusions}}

\vspace{3mm}

The role of the constant $\Lambda$ in cosmology has been made by numerous researchers  a pivotal point of their investigations.
At the same time, it is important to study the effects of $\Lambda$ on an individual particle, in particular in the context of reaction of highly relativistic fermions on gravitational fields \cite{Pl15a}. In the classical (nonquantum) approximation an appropriate instrument for studying the spin-gravity coupling are the MP equations. 

In the  presented paper we study the effects of the dynamic equilibrium (hovering) of the spinning particle relative to the gravitational source of the Schwarzschild--de Sitter background using the MP equations. The analytical solutions of these equations that describe the nonequatorial circular orbits are obtained. It is shown that such orbits exist in both  regions $r<3M$ and $r>3M$, but at the fixed orientation of the spin of the particle the directions 
of its orbital motion are different. For all possible nonequatorial circular orbits in the region $r<3M$ the velocity of the particle is highly relativistic. The similar property takes place in the region $r>3M$, apart from the partial case of the orbits with a small value of $\sin\theta$.

The energy of the spinning particle on  highly relativistic nonequatorial circular orbits is proportional to $\gamma^{-1}$ while the energy of the spinless particle which moves with a highly relativistic velocity along any arbitrary  orbit according to the geodesic equations is proportional to $\gamma$. It means that the contribution of the spin-gravity coupling to the energy of the spinning particle is great.

The obtained data concerning the highly relativistic circular orbits in 
the Schwarz\-schild--de Sitter background provide a good starting point for further analysis of more complicated types of motions of spinning particles,  which can be used to build effective models of astrophysical processes  involving highly relativistic particles.

\end{document}